# Fixed point actions for SU(3) gauge theory [1]

Thomas DeGrand, Anna Hasenfratz
Department of Physics
University of Colorado, Boulder CO 80309-390

Peter Hasenfratz, Ferenc Niedermayer[2]
Institute for Theoretical Physics
University of Bern
Sidlerstrasse 5, CH-3012 Bern, Switzerland

August 1995

**Abstract**

We summarize our recent work on the construction and properties of fixed point (FP) actions for lattice $SU(3)$ pure gauge theory. These actions have scale invariant instanton solutions and their spectrum is exact through 1–loop, i.e. in their physical predictions there are no $a^n$ nor $g^2 a^n$ cut–off effects for any $n$. We present a few-parameter approximation to a classical FP action which is valid for short correlation lengths. We perform a scaling test of the action by computing the quantity $G = L\sqrt{\sigma(L)}$, where the string tension $\sigma(L)$ is measured from the torelon mass $\mu = L\sigma(L)$, on lattices of fixed physical volume and varying lattice spacing $a$. While the Wilson action shows scaling violations of about ten per cent, the approximate fixed point action scales within the statistical errors for $1/2 \geq aT_c$.

[1] Work supported in part by Schweizerischer Nationalfonds, NSF Grant PHY-9023257 and U. S. Department of Energy grant DE–FG02–92ER–40672
[2] On leave from the Institute of Theoretical Physics, Eötvös University, Budapest

Most of the nonperturbative studies of quantum field theory used today involve introducing a discrete space–time lattice as an ultraviolet regulator. Cut–off independent continuum quantities are obtained in the limit of zero lattice spacing. However the lattice introduces artifacts like rotational symmetry breaking. The main problem for numerical lattice calculations is the control of finite lattice spacing effects.

The influence of lattice artifacts on numerical results can depend strongly on the specific regularization or lattice action. Most numerical calculations of QCD to date use the Wilson plaquette action. This action is the simplest one which can be used but simulations done with it do not show scaling unless the lattice spacing is less then 0.1 fm and, consequently, nowadays typical pure gauge or quenched calculations are done on very large spatial volumes[1].

Two of us [2] proposed using a lattice action which is completely free of lattice artifacts. That such a "perfect" action exists follows from Wilson's renormalization group theory [3]. The fixed point (FP) of a renormalization group (RG) transformation and the renormalized trajectory (RT) form a perfect quantum action. The FP action itself reproduces all the important properties of the continuum classical action (it is a classical perfect action). Although at finite coupling the FP action is not "quantum perfect" it is expected to be a very good first approximation. For the 2–dimensional non–linear $\sigma$ model an action parametrized by about 20 parameters was found that showed no finite lattice spacing effects even at correlation length $\xi \sim 3$ lattice spacings, at a negligible computational overhead compared to the gain obtained by the great reduction of lattice artifacts.

This note is a summary of work we have presented in two papers [4, 5] constructing and testing a FP action for lattice $SU(3)$ pure gauge theory. The goal of this program is to find lattice actions which can be used in simulations and which show improved scaling properties compared to the Wilson action. Along the way we discovered some remarkable properties of FP actions, whose relevance may extend beyond lattice calculations.

The partition function of an SU(N) gauge theory defined on a hypercubic lattice is

$$Z = \int DU e^{-\beta S(U)}, \qquad (1)$$

where $\beta S(U)$ is some lattice representation of the continuum action. It is a function of the products of link variables $U_\mu(n) = e^{iA_\mu(n)} \in$ SU(N) along arbitrary closed loops. The normalization is fixed such that on smooth configurations the action takes the standard continuum form $\beta S(U) \to \frac{1}{2g^2} \int d^4x \operatorname{Tr}(F_{\mu\nu}F_{\mu\nu})$, where $\beta = 2N/g^2$.

Let us denote the couplings of $\beta S(U)$ by $\beta$, $c_2, c_3, \ldots$. The action can be represented by a point in the infinite dimensional coupling constant space. A RG transformation introduces blocked links on a coarse lattice and integrates out the original lattice variables, thus increasing the lattice spacing while leaving the physical long distance predictions unchanged. The blocked link variable $V_\mu(n_B) \in$ SU(N) is coupled to a local average of the original link variables. The new action is defined as

$$e^{-\beta' S'(V)} = \int DU \exp\left\{-\beta\left(S(U) + T(U,V)\right)\right\}, \tag{2}$$

where $T(U,V)$ is the blocking kernel.

Under repeated real space renormalization group transformations the action moves in this multidimensional space. On the critical surface it flows to a fixed point. For an asymptotically free theory the critical surface is at $\beta \to \infty$. In this limit Eq. (2) reduces to a saddle point problem giving

$$S^{FP}(V) = \min_{\{U\}} \left(S^{FP}(U) + T(U,V)\right), \tag{3}$$

where $S^{FP}$ is the FP action. The FP action is constructed by solving Eq. (3). This action is classically perfect: its spectrum exhibits no lattice artifacts. It also has scale invariant instanton solutions [6].

In addition to being classically perfect, the FP action is also one-loop perfect. Moving away from the fixed point along the RT, the only effect of the perturbative $g^2$ corrections is to make the marginal operator $S^{FP}(U)$ weakly relevant, i.e. the coupling $\beta = 2N/g^2$ begins to move, but the form of the action on the renormalized trajectory remains unchanged through $O(g^2)$:

$$\beta S^{FP}(U) \longrightarrow \beta' S^{FP}(V), \tag{4}$$

under a RG transformation, to 1–loop level. Here $\beta' = \beta - \Delta\beta$, where $\Delta\beta$ is fixed by the first universal coefficient of the $\beta$–function. An equation analogous to Eq. (4) was presented in ref. [7] without discussion.

The formal RG argument is very simple. Parameterize the theory in terms of one marginal operator $R_1$, its coupling $d_1$, and many irrelevant operators $R_n(U)$, $n = 2, \ldots$ and their associated couplings $d_n$. Under an RG transformation

$$d'_k = \lambda_k d_k + \alpha^{(k)}_{ij} d_i d_j + O(d^3). \tag{5}$$

where $\lambda_1 = 1$ and $\lambda_n < 1$ for $n > 1$. If we start with the action

$$S^{FP} = R_1, \tag{6}$$

$d_1$ small, $d_2 = d_3 = \ldots = 0$, i. e. the classical fixed point action, then after one RG step

$$d'_1 = d_1 + \alpha_{11}^{(1)} d_1^2 + O(d^3) \qquad (7)$$
$$d'_k = \alpha_{11}^{(k)} d_1^2 + O(d^3), \qquad (8)$$

or

$$\frac{1}{g^2} S^{FP} \to \frac{1}{g^2 + \alpha_{11}^{(1)} g^4} (S^{FP} + \alpha_{11}^{(2)} g^4 R_2 + \alpha_{11}^{(3)} g^4 R_3 + \ldots) \qquad (9)$$

so that the action itself is changed only in order $g^4$ to all orders in the latticed spacing $a$. Note that variations on the Symanzik[8, 9] program yield actions which are only $O(a^2)$ or $O(g^2 a^2)$ improved.

An explicit perturbative calculation with the FP action of the non–linear $\sigma$–model in $d = 2$ [10] confirms this picture. In ref. [10] the mass gap $m(L)$ was calculated in a finite periodic box of size $L$. The prediction of the standard action is contaminated by power like cut–off effects $(a^2/L^2)^n, n = 1, 2, \ldots$. With the FP action these type of cut–off effects are reduced to the level of the numerical precision (dominated by the uncertainties in the couplings of the quartic part of the action). The coefficient of the leading $a^2/L^2$ cut–off effect is reduced by 4 orders of magnitude relative to that of the standard action.

The action on the renormalized trajectory is perfect — it has only physical excitations. In a Green's function only physical states come in as intermediate states independent of the operator used. In spite of that Green's functions (of fields, currents, etc.) generally show cut–off effects. Consider, for example a free scalar field and imagine constructing the FP action by making many blocking steps, beginning on a very fine lattice. The FP lattice action is described by a lattice field which is the average of the continuum field over a hypercube. Therefore, the two-point function of the lattice field in the FP action is equal to the two-point function of the continuum field averaged over the hypercubes around the the two lattice points in question. This averaging brings in a trivial rotation symmetry breaking. The two-point function of the field in the FP action is not rotationally invariant, although only physical states propagate. The same result is true for the potential derived from Polyakov loop correlators: the naive correlator has power law corrections to the true potential.

These power law corrections can be eliminated by constructing fixed point operators — operators which the naive operators block into under repeated RG transformations. We have constructed the FP Polyakov loop for smooth gauge configurations (quadratic action) and used its correlator to compute the perturbative $\beta \to \infty$ potential $V(r)$. We display the classical potential from the FP action using its FP operator, in Fig. 1 and contrast it to the potential computed from the Wilson action using the naive operator, shown in Fig.

2. One can prove that the FP correlator has no power law corrections, only exponentially vanishing $\simeq \exp(-r/a)$ ones.

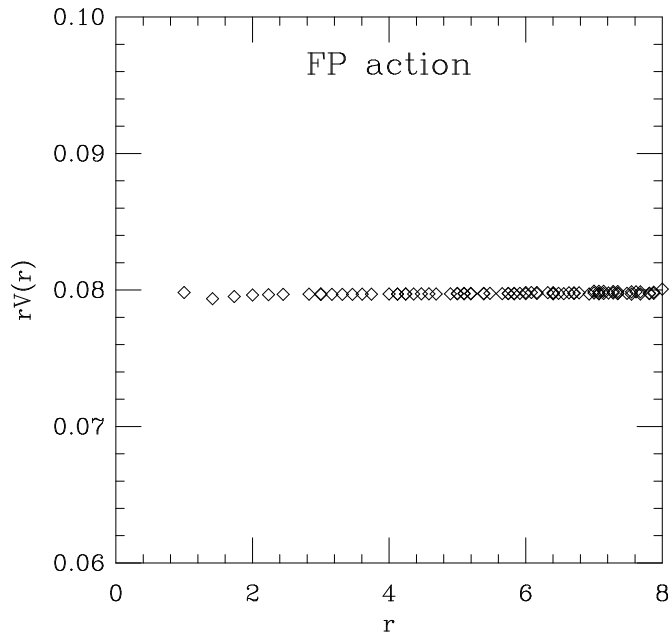

Figure 1: The quantity $rV(r)$ constructed from the correlators of the FP Polyakov loops and using the FP action. In the continuum $rV(r) = 1/4\pi = .0796$.

We have constructed FP actions for two types of RG transformations. The one we have studied most extensively is based on a scale factor 2 blocking [11]. It has tunable parameters, which we adjust to minimize the range of the action. We did this construction in two steps. First, we solved the RG equations for smooth gauge configurations. In this limit the action can be taken to be an arbitrary quadratic function of the vector potential. The solution of the RG equations can be formulated analytically and the computer is only used to do Fourier mode sums. Here we do the tuning of parameters for locality. To go beyond leading order we solved the fixed point equation Eq. (3) numerically. The procedure is the following: We create an arbitrary SU(3) configuration {V}. That serves as a coarse configuration. We then minimize the RHS of Eq. (3) numerically. As an intermediate step we approximate $S^{FP}$ with a simpler action $S_0$. We choose $S_0$ such that it is sufficiently simple for minimization but close to

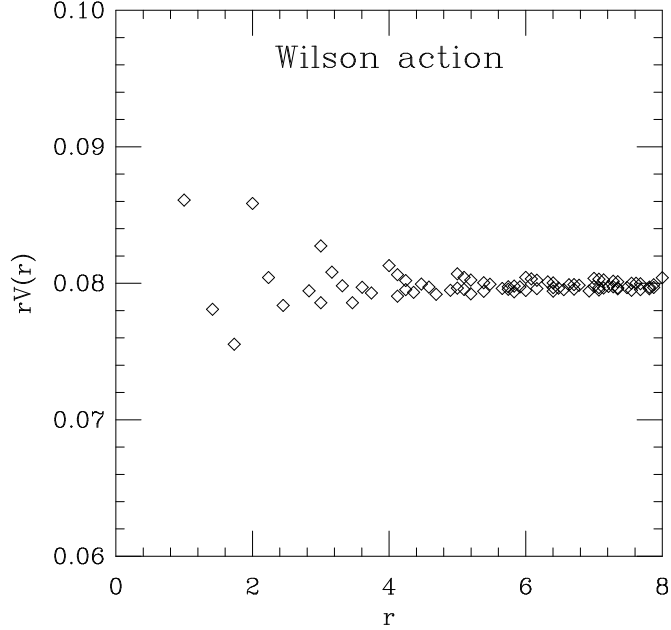

Figure 2: The quantity $rV(r)$ constructed for the Wilson action.

the fixed point action. The minimization of Eq. (3) gives the fine configuration $\{U_0\}$ for this $\{V\}$

$$S'(V) = \min_U (S_0(U) + T(U,V)) \qquad (10)$$
$$= S_0(U_0) + T(U_0, V).$$

If $S_0$ is close to the fixed point action $S^{FP}$, than $U_0$ is close to $\{U^{FP}\}$ and up to quadratic corrections

$$S^{FP}(V) = S^{FP}(U_0) + T(U_0, V)$$
$$= S^{FP}(U_0) + (S'(V) - S_0(U_0)). \qquad (11)$$

This equation can be used to calculate the value of the fixed point action on the course configuration $\{V\}$ if $S^{FP}$ is known on the fine configuration $\{U_0\}$.

We generated about 400 configurations on $2^4$ and $4^4$ coarse lattices using the Wilson action, with coupling $\beta$ ranging from 5.0 to 7.0 and solved Eq. (11), parameterizing the FP action as a sum of powers of traces of loops. The fitting procedure has considerable uncertainty in it: we are attempting to construct a practical approximation to a FP action subject to the constraint that it be easy to use. It is not obvious to us what set of coarse configurations constitutes a "reasonable" data set to use in the construction, nor how closely the action must approximate a true FP action in order to perform well in simulations. These problems are open questions.

For numerical simulation we need a parametrization which is simple and approximates the FP action sufficiently well on those configurations which are typical in the actual calculation. In order to find such a parametrization we fitted $S^{FP}(V)$ with powers of the traces of only 2 loops. We took the plaquette and the twisted 6 link operator (x,y,z,-x,-y,-z). We kept up to four powers of $(3 - \text{Re} \text{Tr} U)^n$. The 8 parameters are listed in Table 1.

Table 1: Couplings of the few-parameter FP action for the RG transformation.

| operator | $c_1$ | $c_2$ | $c_3$ | $c_4$ |
|---|---|---|---|---|
| $c_{plaq}$ | .523 | .0021 | .0053 | .0167 |
| $c_{6-link}$ | .0597 | .0054 | .0051 | -.0006 |

We simulated this action with a mixed Metropolis-overrelaxation algorithm acting on SU(2) subgroups, in complete analogy with a standard Wilson action code. No special optimization was done. The program runs about a factor of 5-7 slower than a highly optimized Wilson code.

Scaling means that all physical dimensional quantities show the same functional dependence on the gauge coupling. It is different from asymptotic scaling where in addition to scaling we require that this functional dependence be described by the 2-loop $\Lambda$ parameter. Asymptotic scaling properties can be significantly improved by a suitable non-perturbative redefinition of the bare coupling [12, 13]. However, here we are not concerned with asymptotic scaling. Scaling properties can only be changed by modifying the lattice action.

We used the critical temperature $T_c$ to set the physical scale for our scaling test. We determined the critical coupling constant $\beta_c(N_t)$ for $N_t = 2, 3, 4$ and 6 and fixed the lattice spacing at these coupling values as $a = 1/(T_c N_t)$.

Next we measured physical observables at couplings $\beta_c(N_t)$ in fixed, finite physical volumes. This way we can avoid infinite volume extrapolations. For one scaling test we consider the quantity $G = L\sqrt{\sigma(L)}$ where $\sigma(L)$ is the string tension on an $L^3$ volume computed from the exponential fall-off of the Polyakov line correlator (torelon mass). As we vary the lattice coupling we also vary the lattice spatial size so that the physical volume is kept fixed at $V = (r/T_c)^3$, where $r$ is some conveniently chosen aspect ratio. The quantity $G$ is indepen-dent of the bare coupling (resolution) in the scaling (continuum) limit. Any variation of $G$ is due to lattice artifacts. Figure 3 shows $G(L)$ with $r = 2.0$ as a function of $aT_c$ for the Wilson and for the 8 parameter FP action. The inner and outer error bars on the FP points show their statistical uncertainty, and the combined uncertainty from statistics and in $\beta_c$. The Wilson action shows a scaling violation of about 10% between $N_t = 2$ and 6. No scaling violation above the statistical errors is seen for the FP action.

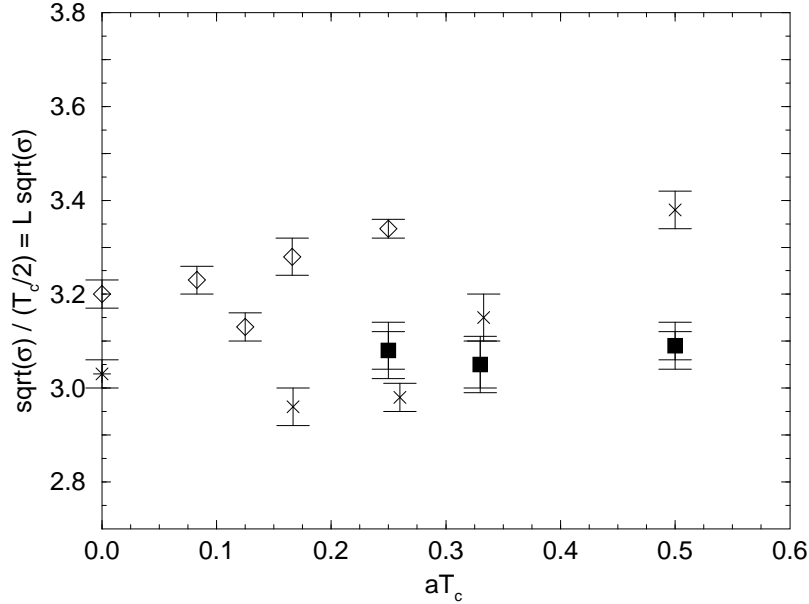

Figure 3: Scaling test for the Wilson action (crosses) and FP action (squares); $T_c$ is defined in infinite spatial volumes. The diamonds are from zero-temperature simulations by Ref. 14 and the diamond and burst at $a = 0$ are extrapolations.

No published simulations with the Wilson action at small lattice spacing precisely match our scaling test using the torelon mass on fixed physical volumes. The ones which come closest are those of Boyd et. al. [14]. They present measurements of the quantity $\sqrt{\sigma}/T_c$ at the critical couplings for $N_t = 4$, 6, 8, and 12, for which the string tension has been computed via fits to Wilson loops on large (zero temperature) lattices. The string tension from Wilson loops is an upper bound on $\sigma(L)$; $\sigma(L)$ is reduced from $\sigma$ by the zero-point fluctuation term which for large $L$ in a string model is

$$\sigma(L) = \sigma - \frac{\pi}{3L^2} + \ldots . \tag{12}$$

The diamonds in Fig. 3 show $2\sqrt{\sigma}/T_c$ from Ref. [14]. The authors of that work present an extrapolation to $a = 0$, which we also display along with the $a = 0$ limit of $L\sqrt{\sigma(L)}$ at $r = 2$ calculated from Eq. (12). The FP action data scales within its uncertainty and appears to agree with the zero-lattice spacing extrapolation.

We have described a systematic program for constructing fixed point actions for SU(3) gauge theory, and illustrated it using a particular scale two blocking transformation. An approximation to the FP action shows scaling within our small statistical errors beginning at $aT_c = 1/2$, as compared to $aT_c \leq 1/8$ for the Wilson action, at a cost of a factor of about 5-7 in computational speed per site update. The outstanding problem now is to extend the program to include fermions. The free fermion case, the Schwinger model and the large $N$ limit of the Gross–Neveu model have been considered in this context [15].

# 1   Acknowledgements

U. Wiese participated in the early stages of this project. We are indebted to M. Blatter, R. Burkhalter, P. Kunszt and P. Weisz for valuable discussions. We would like to thank T. Barker, M. Horanyi and the Colorado high energy experimental group for allowing us to use their work stations. We want to thank P. Büttiker, M. Egger and M. Willers for their support in using the computing facilities at the University of Bern. This work was supported by the U.S. Department of Energy and by the National Science Foundation and by the Swiss National Science Foundation.

# References

[1] Cf. the summary talks of D. Weingarten, Lattice '93, Nucl. Phys. **B** (Proc. Suppl.) 34 (1994) 453 and C. Michael, Lattice '94, Nucl. Phys. **B** (Proc. Suppl.) 42 (1995) 147.

[2] P. Hasenfratz and F. Niedermayer, Nucl. Phys. B414 (1994) 785; P. Hasenfratz, Nucl. Phys. B (Proc. Suppl) 34 (1994) 3; F. Niedermayer, *ibid.*, 513.

[3] K. Wilson and J. Kogut, Phys. Rep. C12 (1974) 75;
K. Wilson, Rev. Mod. Phys. 47 (1975) 773, *ibid.*, 55 (1983) 583.

[4] T. DeGrand, A. Hasenfratz, P. Hasenfratz, F. Niedermayer, preprint COLO-HEP-361, BUTP–95/14.

[5] T. DeGrand, A. Hasenfratz, P. Hasenfratz, F. Niedermayer, preprint COLO-HEP-362, BUTP–95/15 (1995).

[6] For a discussion of instantons in the $O(3)$ spin model, see M. Blatter, R. Burkhalter, P. Hasenfratz and F. Niedermayer, Nucl. Phys. **B** (Proc. Suppl.) 42 (1995) 799 and preprint BUTP-95/17.

[7] K. Wilson, *in* Recent developments of gauge theories, ed. G. 't Hooft et al. (Plenum, New York, 1980).

[8] K. Symanzik, in "Recent Developments in Gauge Theories," eds. G. 't Hooft, et. al. (Plenum, New York, 1980) 313; in "Mathematical Problems in Theoretical Physics," eds. R. Schrader et. al. (Springer, New York, 1982); Nucl. Phys. B226 (1983) 187, 205.

[9] P. Weisz, Nucl. Phys. B212 (1983) 1. M. Lüscher and P. Weisz, Nucl. Phys. B240[FS12] (1984) 349; G. Parisi, Nucl. Phys. B254 (1985) 58; M. Lüscher and P. Weisz, Comm. Math. Phys. 97 (1985) 59.

[10] A. Farchioni, P. Hasenfratz, F. Niedermayer and A. Papa, BUTP-95/16, IFUP-TH 33/95, preprint (1995).

[11] R. H. Swendsen, Phys. Rev. Lett. 52 (1984) 2321; A. Hasenfratz, P. Hasenfratz, U. Heller, and F. Karsch, Phys. Lett. 140B (1984), 76; K. C. Bowler, et. al., Nucl. Phys. B257[FS14] (1985), 155 and Phys. Lett. 179B (1986) 375; QCD-TARO collaboration (K. Akemi, et. al.), Phys. Rev. Letters 71 (1993), 3063.

[12] G. Parisi, in the Proceedings of the XX International Conference on High Energy Physics, L. Durand and L. Pondrom, eds., American Institute of Physics, p. 1531.

[13] G. P. LePage and P. Mackenzie, Phys. Rev. **D48** (1993) 2250.

[14] G. Boyd, et. al., Bielefeld preprint BI-TP 95/23, June 1995.

[15] U.-J. Wiese, Phys. Lett. B315 (1993) 417 , W. Bietenholz, E. Focht and U.-J. Wiese, Nucl. Phys. B436 (1995) 385, W. Bietenholz and U.-J. Wiese, MIT preprint, CTP 2423 (1995).